\documentclass[aps,showpacs,amsmath,amssymb]{revtex4}

\usepackage{graphicx}
\usepackage{dcolumn}
\usepackage{bm}
\usepackage{natbib} 
\newcommand{\um}{$\mu$m}
\newcommand{\puw}{$P_{\mu w}$}
\newcommand{\mpuw}{P_{\mu w}}

\begin{document}

\title{Temperature Dependence of the Frequency and Noise of Superconducting Coplanar Waveguide Resonators}

\author{Shwetank Kumar, Jiansong Gao, Jonas Zmuidzinas}
\affiliation{Division of Physics, Mathematics, and Astronomy\\
California Institute of Technology, 
Pasadena, CA 91125}
\author{Benjamin A. Mazin, Henry G. LeDuc, Peter K. Day} 
\affiliation{Jet Propulsion Laboratory, California Institute of 
Technology, Pasadena, CA 91109}

\date{\today}

\begin{abstract}
We present measurements of the temperature and power dependence
  of the resonance frequency and frequency noise of
  superconducting niobium thin-film coplanar waveguide resonators,
  carried out at temperatures well below the superconducting
  transition ($T_c = 9.2$\ K).
  The noise decreases by nearly two orders of magnitude as the temperature
  is increased from 120 to 1200~mK, while
  the variation of the resonance frequency with temperature over this range
  agrees well with the standard two-level
  system (TLS) model for amorphous dielectrics. These results
  support the hypothesis that TLS are responsible for the noise in
  superconducting microresonators, and have important implications for
  resonator applications such as qubits and photon detectors.
\end{abstract}


\maketitle

Superconducting microresonators are useful for photon
detection\cite{Mazin02,Day03,Schmidt05,Mazin06,
Barends07,Baselmans08,Vardulakis08}, coupling to
qubits\cite{Wallraff04,Lee05,Sillanpaa07}, SQUID multiplexers
\cite{Irwin04,Lehnert07}, and for studying basic physics
\cite{Abdo07,Yurke06,Abrelsegev07,Tholn07}.  
Such resonators have excess
noise\cite{Day03,Mazinthesis} that is equivalent to a jitter of the
resonance frequency,
likely caused by two--level tunneling
systems (TLS) in amorphous dielectrics\cite{Gao07}.
Indeed, TLS models explain the unusual
properties of amorphous materials at low temperatures
\cite{Hunklinger72,Schickfus75, Laikhtman85,Phillips87}, and recent
qubit experiments\cite{Martinis05} have highlighted 
TLS effects in superconducting microcircuits.
While the TLS energy splitting $\Delta E$ has a broad
distribution\cite{Phillips87},
a resonator with frequency $f_r$ is most sensitive 
to TLS with $\Delta E \sim h f_r$.
The level populations and relaxation rates of such TLS
vary strongly at temperatures $T \sim h f_r /2k_B$,
or around 100~mK for the device studied here. 
Furthermore, such near-resonant TLS may saturate\cite{Gao07} for 
strong resonator excitation power \puw.  
Hence, measurements of the power and temperature variation of 
the resonator frequency and noise, as presented in this
letter, provide a strong test of the TLS hypothesis.  

We studied coplanar waveguide (CPW)\ quarter--wavelength resonators\cite{Day03,Gao07} 
fabricated on a high-resistivity ($\rho \geq 10$~k$\Omega$\,cm) crystalline silicon
substrate by patterning a 200 nm thick niobium film using a photoresist mask and an SF$_6$
inductively--coupled plasma etch.
In this device, TLS may be present in the native oxide surface
layers on the  metal film or substrate\cite{Gao07}.
The resonator is capacitively coupled to a CPW feedline (Fig.~1) that has a
10\,\um\ wide center strip and 6\,\um\ gaps between center strip and
the ground plane.  For the resonator, these dimensions are 5\,\um\ and
1\,\um\ respectively.  The resonator length is 5.8~mm, corresponding
to $f_r = 4.35$~GHz. 
The coupling strength is set lithographically\cite{Mazinthesis,Gao07}
(see Fig. 1) and is characterized by the coupling--limited quality 
factor $Q_c = 5 \times 10^5$.


The device was cooled using a dilution refrigerator, and its temperature was
measured to $\pm 5$~mK accuracy using a calibrated
RuO$_{2}$ thermometer mounted on the copper sample enclosure.
The microwave readout (Fig.~1) uses a standard $IQ$ homodyne mixing
technique\cite{Day03,Mazinthesis}.
The $IQ$ mixer's complex output voltage $\xi(f) = I(f) + j Q(f)$
follows a circular trajectory  in the complex plane as the microwave
excitation frequency $f$ is varied\cite{Gao07},
and $f_r$ and $Q_r$ are determined by complex least--squares fitting 
of this trajectory to a ten--parameter model:

\begin{equation}
Z^{(model)}(f) = \left( B_{0}+B_{1} \delta x \right) \, 
                 \exp[i(\phi _{0}+\phi _{1} \delta x)] \,
                \left[\frac{S_{21}^{(r)}+2jQ_r\delta x}
                           {1+2jQ_r\delta x}\right]
              + B_{2}\, \exp[i\phi_{2}]\ .
\label{eq:lorentzian}
\end{equation}
Here $\delta x = (f-f_{r})/f_{r}$ is the fractional frequency offset,
$S_{21}^{(r)}$ is the complex forward transmission on resonance,
$B_0 + B_1 \delta x$ allows for a linear gain variation,
$\phi _0 + \phi_1 \delta x$ allows a similar linear phase variation,
and $B_{2}$ and $\phi _{2}$ specify the output offset voltages
of the $IQ$ mixer.

The combined noise of
the resonator and readout electronics is measured by tuning the
synthesizer to the
resonance ($f = f_r$) and digitizing the fluctuations 
$\delta \xi(t)$\cite{Day03,Mazinthesis,Gao07}.  
The noise analysis follows our previous work\cite{Gao07}.  
In brief, the noise covariance matrix is
calculated and diagonalized at each noise frequency,
yielding power spectra for amplitude (dissipation) and phase (frequency) fluctuations.
The amplitude noise is consistent with the 
electronics noise floor measured off resonance, so we
estimate the resonator's frequency noise by subtracting
the amplitude noise spectrum from the phase noise spectrum.

The quality factor $Q_r$ lies in the range
$1.5\times 10^5$ to $4.5 \times 10^5$ and is both temperature
and power dependent; however, the interpretation
is not straightforward due to TLS saturation 
effects\cite{Martinis05,Gao07,Martinis07}.
In contrast, $f_r$ is much less affected by
TLS saturation\cite{Martinis07}.
Figure 2a shows the measured temperature dependence of the frequency shift
$\delta f_r(T,\mpuw)$. Note that $\delta f_r$ increases with temperature,
whereas the Mattis--Bardeen theory\cite{Mattis58} 
predicts a superconductivity-related frequency shift 
that is much smaller and opposite in sign.  
Instead, the data fit quite well to the functional form predicted 
TLS theory\cite{Phillips87}.  Above 900~mK, this fit can be further
improved by including the Mattis-Bardeen contribution, according to
the following model:
\begin{eqnarray}
\frac{\delta f_r^{(model)}(T, \mpuw)}{f_r} &=& C_1(\mpuw) 
              + C_2(\mpuw) \left[ \mathrm{Re} \, 
                \psi\left(\frac{1}{2}+\frac{hf_r}{2\pi i k_bT}\right) -
                \log \left(\frac{hf_r}{k_bT}\right) \right] \nonumber \\
            && +  \frac{C_3}{4} \left[
                \frac{\sigma_2(T)-\sigma_2(0)}{\sigma_2(0)} \right]\ .
\label{eqn:digamma}
\end{eqnarray}
There are three free parameters for each power level:
$C_1(\mpuw)$ allows a small power--dependent frequency shift
relative to $f_r(120\,{\rm mK}, -72\,{\rm dBm})$;
$C_2(\mpuw)$ is the coefficient of the TLS linear response term\cite{Mason76} 
and is allowed to vary with power to account for possible TLS saturation;
$C_3$ is the kinetic inductance fraction of the CPW line\cite{Gao06b}
and should be constant over the range of readout power used.
The data were fit for readout power values from -72~dBm to -92~dBm in
steps of 4~dBm.  The value of $C_3$ was indeed found to be constant
($C_3 = 0.104 \pm 0.021 $) and in close agreement with the expected
value\cite{Gao06b} $C_3 = 0.125$.  Meanwhile,
$C_1 = 1.948 \pm 0.002 \times 10^{-5}$ at -72~dBm
and 
$1.902\pm0.002 \times 10^{-5}$ at -92~dBm,
while 
$C_2 = 9.09 \pm 0.02 \times 10^{-6}$
and 
$9.39 \pm 0.02 \times 10^{-6}$
for -72 and -92~dBm, respectively.
For a fixed temperature $T$, we find that the frequency shift
$\delta f_r$ scales with power approximately as $P_{\rm int}^{0.3}$,
where $P_{\rm int} = 2 Q_r^2 \mpuw / (\pi Q_c)$ is the resonator's
internal  microwave power.

The coefficient $C_2$ is a measure of
the number of TLS that are coupled to the resonator's electric field
$\vec E(\vec r)$.  This relationship may be quantified in terms of the
microwave loss tangent $\delta$ of the amorphous TLS material and the
$|\vec E|^2$--weighted volume filling fraction $F$ of that
material, according to $C_2 = F \delta/\pi$.  Typical amorphous
materials have loss tangents of order $\delta \sim 10^{-2} - 10^{-3}$.
Assuming a uniform distribution of TLS on the surface of the
resonator, perhaps due to surface oxides, a reasonable oxide thickness
of order 10\,nm would be consistent with the observed filling
factor $F \sim 10^{-2}$.

The temperature and power dependence of the resonator noise was
quantified by first calculating the fractional frequency noise
spectrum\cite{Gao07}, $S_{\delta f_r}(\nu)/{f_r}^2$, which was then
averaged over the range 200-300~Hz, a clean portion of the spectrum
well above the HEMT noise floor at low temperatures.  The resulting
values are plotted in Fig.~3, demonstrating the very strong
temperature dependence of noise.
These data may be described reasonably well by the fitting function
\begin{equation}
\frac{\bar{S}_{\delta f_r}(\nu)}{f_r^2} = A \mpuw^{\alpha} T^{\beta}
\tanh^2 \left( \frac{h f_r}{2 k_B T}\right)
\end{equation} 
with indices $\alpha = -0.46 \pm 0.005$ and $\beta = -0.14 \pm 0.02$.
The value of $\alpha$ is consistent with previous work \cite{Gao07}.
While the inclusion of the hyperbolic tangent factor is motivated by a 
particular model for the TLS noise\cite{Martinis07}, and the
low value of $\beta$  is suggestive, at present the data cannot
distinguish between this model and a simple power law temperature
dependence of the form $T^{-1.73}$. Additional measurements,
especially at lower temperatures, will be needed to further
elucidate the physical mechanism of the TLS noise.

In conclusion, both the frequency and noise of our superconducting
CPW resonators show substantial variation at temperatures far below
the superconducting critical temperature. The variation of the resonance
frequency is well described by TLS theory with plausible values for
the loss tangent and filling factor. Combined, these results strongly
suggest that the resonator noise is also due to TLS and is not related
to the superconductor.  The temperature dependence of the noise also
has important practical implications. For instance, if the TLS origin
of the noise is correct, designing resonators to operate in the regime
$f_r << 2kT/h$ could result in lower noise and improved performance.

We thank Sunil Golwala, Kent Irwin, Andrew Lange, Konrad Lehnert, 
and Harvey Moseley, and especially John Martinis for useful discussions.
This work was
supported in part by the NASA Science Mission Directorate, JPL, 
and the Gordon and Betty Moore Foundation.

\newpage

\noindent FIGURE CAPTIONS
\\

\noindent Figure 1\\
The experimental setup is illustrated.
The resonator (not to scale) is shown schematically;
black represents the superconducting film; white represents
bare substrate.
The resonator is excited using a microwave synthesizer,
and its output signal is sent to a cooled HEMT
amplifier  with $\sim 4$\,K noise temperature followed 
by a room-temperature amplifier.
Amplitude and phase information are recovered simultaneously
using an $IQ$ mixer.
The attenuators $A_1$ and $A_2$ allow the incident microwave power \puw\ 
to be varied over a wide range while maintaining the optimal
power level at the $IQ$ mixer's input by constraining the sum of the 
attenuations $A_1+A_2$ to be constant (in dB).
The $IQ$ output voltages
are amplified, digitized, and recorded with 16-bit resolution at 250~kSa/s.
\\

\noindent
Figure 2\\
(a) The resonance frequency shift, defined as
$\delta f_r(T,\mpuw) = f_r(T,\mpuw) - f_r(120\,{\rm mK},-72\,{\rm dBm})$,
 is plotted as a function of temperature for readout powers of -72 dBm
(filled circle) and -92 dBm (filled square). The dashed
line shows the frequency shift predicted by the Mattis-Bardeen
theory, but scaled up by a factor of 100.  The solid lines represent fits
to the data using equation \ref{eqn:digamma}.  (b) A plot of the
residuals after subtracting the fit from the data; the representative
error bar indicates that the fit matches the data to within the
$\pm 5$~mK accuracy of the thermometry.
\\

\noindent
Figure 3\\
The average value of the fractional frequency noise power spectrum 
in the 200--300~Hz range ($\bar{S}_{\delta f_r}(\nu)/f_r^2$)
is plotted as a function of temperature ($T$) for several values of
the microwave readout power (\puw).
The power levels \puw range from -96~dBm to -72~dBm, as indicated
by the labels on the right.


\newcommand{\placefigures}{
\newpage
\begin{figure}
  \begin{center}
  \resizebox{6.5in}{!}{\includegraphics{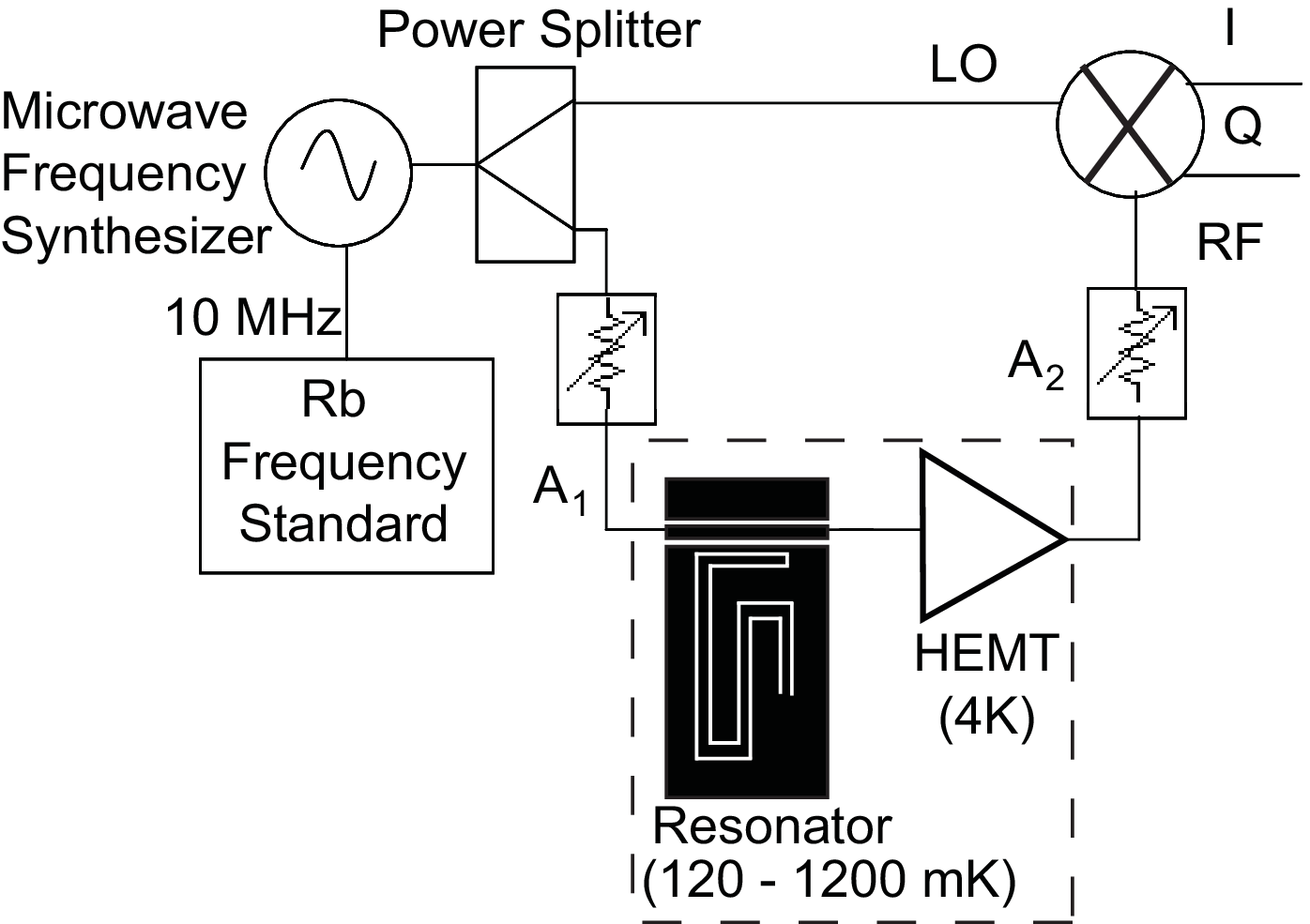}}\\
\vspace{1in}
Figure 1
  \end{center}
\end{figure}
\newpage
\begin{figure}
  \begin{center}
  \resizebox{6in}{!}{\includegraphics{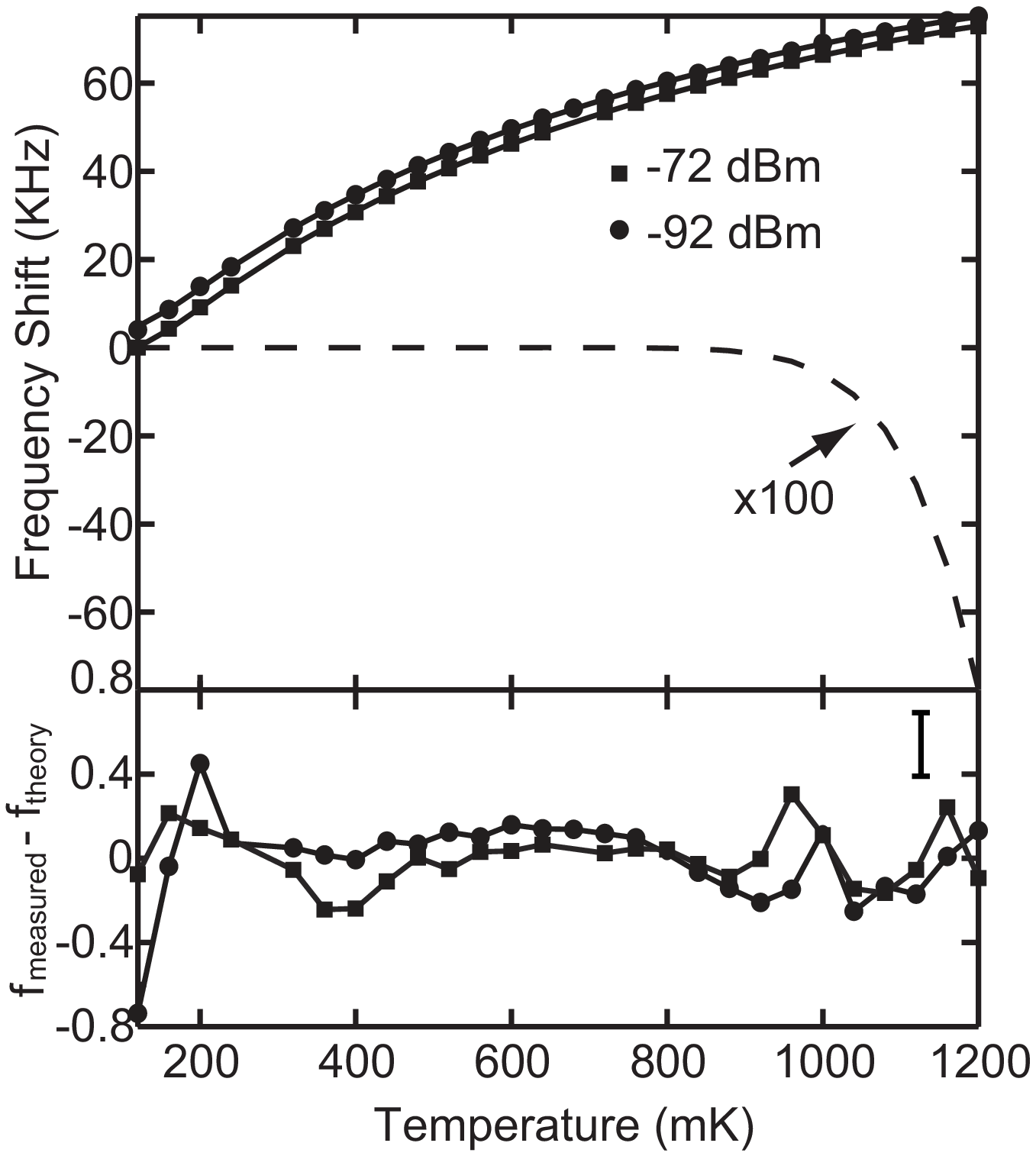}}\\
\vspace{1in}
Figure 2
  \end{center}
\end{figure}
\newpage
\begin{figure}
  \begin{center}
  \resizebox{6in}{!}{\includegraphics{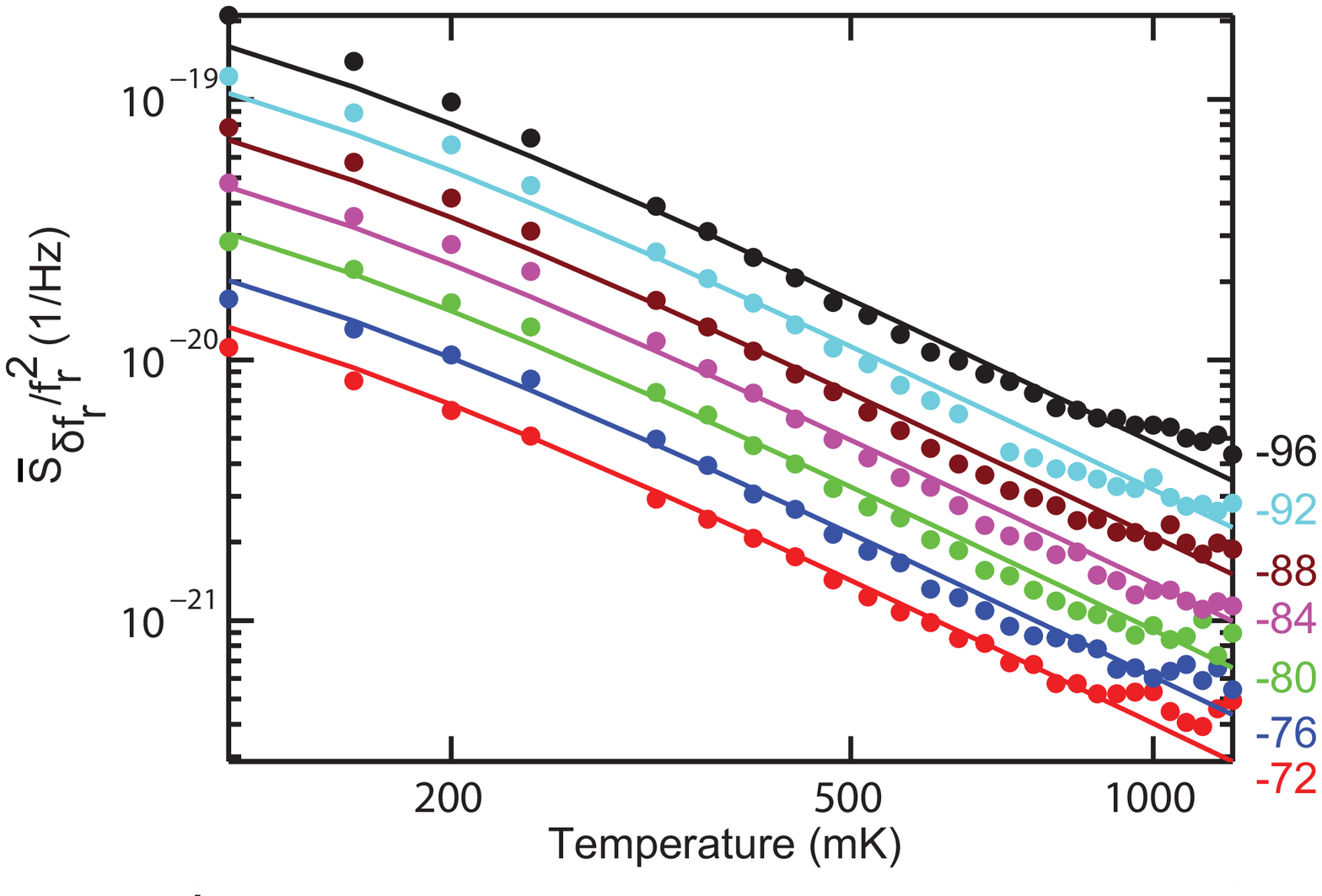}}\\
\vspace{1in}
Figure 3
  \end{center}
\end{figure}
}
\placefigures
\end{document}